\documentclass[11pt]{article}
\usepackage{times}
\usepackage[T1]{fontenc}
\usepackage{fullpage}
\usepackage{latexsym}
\usepackage{amsmath}
\usepackage{amssymb}
\usepackage{amsthm}
\usepackage{epsfig}
\usepackage{fullpage}
\usepackage{verbatim}
\usepackage{color}
\usepackage{graphicx}
\usepackage{todonotes}
\usepackage{array,multicol, multirow,siunitx,tabularx}
\usepackage[noframe]{showframe}
\usepackage{framed}
\usepackage[utf8]{inputenc}
\usepackage{amsthm}
\usepackage{multicol}
\usepackage{graphicx}
\usepackage{caption}
\usepackage{subcaption}

\newtheorem{theorem}{Theorem}[section]

\newtheorem{lemma}[theorem]{Lemma}
\newtheorem{definition}[theorem]{Definition}

\newtheorem{claim}[theorem]{Claim}
\newtheorem{proposition}[theorem]{Proposition}

\newcommand{\stab}{\mathrm{Stab}_{\pi}(n,k)}
\newcommand{\stabsize}{m(n,k)}
\newcommand{\sgraph}{SG_{n,k,\pi}}
\newcommand{\chromaticnumber}{\chi(\sgraph)}
\newcommand{\test}{\phi_{x\rightarrow y}}

\newcommand{\E}{\mathop{\mathbb{E}}}

\title{Hardness of Coloring $2$-Colorable $3$-Uniform Hypergraphs with $\log^{\Omega(1)} n$ Colors}
\author{Anonymous}

\author{Amey Bhangale\thanks{Department of Computer Science and Engineering, University of California, Riverside. Supported by NSF CAREER award 2440882.} \and Krishna Priya KM\thanks{Department of Computer Science and Engineering, University of California, Riverside. Supported by NSF CAREER award 2440882.}\and Simon Rubin-Toles\thanks{Department of Computer Science and Engineering,
University of California, San Diego.}}

\begin{document}

\maketitle

\begin{abstract}
    We show that it is quasi-NP-hard to color a $2$-colorable $3$-uniform hypergraph on $n$ vertices using $O((\log n)^{1/8})$ colors. This exponentially improves the result of Dinur, Regev, and Smyth [Combinatorica'2005], which showed it is quasi-NP-hard to color a $2$-colorable $3$-uniform hypergraph on $n$ vertices using $O((\log \log n)^{1/3})$ colors.
\end{abstract}
\section{Introduction}

A hypergraph $\mathcal{H} = (V,E)$ with vertices $V$ and edges $E \subseteq 2^V$ is called \textit{$3$-uniform} if each edge in $E$ contains exactly $3$ vertices. A valid $\chi$ coloring of a hypergraph $\mathcal{H}$ is a function $f:V \to [\chi]$ such that no edge in $E$ is monochromatic. We will examine the hardness of approximately coloring $2$-colorable $3$-uniform hypergraphs.

The best known polynomial-time algorithm is the semi-definite program of Krivelevich, Nathaniel, and Sudakov \cite{KrivNS01}, which requires $O(n^{1/5})$ colors. On the hardness side, our understanding of this problem saw significant improvement with the advent of the PCP theorem. Most notably, the 2002 result of Dinur, Regev, and Smyth introduced the idea of constructing a layered PCPs out of traditional projection PCPs \cite{DinurRS02}. They showed that, assuming $\text{NP} \nsubseteq \text{DTIME}(2^{\mathrm{poly}(\log n)})$, there is no polynomial-time algorithm for coloring a $2$-colorable $3$-uniform hypergraph with $O(\sqrt[3]{\log \log n})$ colors. Crucially, their construction involved the Kneser graph, whose vertices correspond to the $k$-element subsets of a set of $n$ elements with edges between two vertices if and only if the corresponding sets are disjoint, over the Long-Code of the domain $R$ of each layer of the PCP. They showed that any coloring of a Kneser graph with a small number of colors must contain a `special' color that has a monochromatic edge and colors a certain fraction of the vertices, allowing decoding.

The \cite{DinurRS02} result followed the work of Guruswami, H{\aa}stad, and Sudan \cite{GHS02}, who showed that it is hard to color a $2$-colorable $4$-uniform hypergraph with $\Omega(\frac{\log \log n}{\log \log  \log n})$ colors, assuming P $\neq$ NP. Since then, many have studied the problem of hypergraph coloring, improving bounds on variants of the problem. Most of these results are under assumptions stronger than $\text{P} \neq \text{NP}$ when the hardness factors are superconstant: In 2014, Guruswami et. al.~\cite{GHHSV17} showed that it is hard to color a $3$-colorable $3$-uniform hypergraph with $2^{\Omega(\frac{\log\log n}{\log \log \log n})}$ colors, assuming $\text{NP} \nsubseteq \text{DTIME}(n^{2^{\mathcal{O}(\frac{\log\log n}{\log \log \log n})}})$. Also in 2014, Saket~\cite{S14} showed that it is hard to color a $2$-colorable $4$-uniform hypergraph with $\Omega(\log^{\delta}n)$ colors for $0 < \delta \ll 1$ assuming $\text{NP} \nsubseteq \text{DTIME}(n^{\mathcal{O}(\log\log n)})$. In 2018, Bhangale~\cite{Bhangale18} improved this bound to $\Omega(\frac{\log n}{\log \log n})$ under the same assumptions using methods involving {\em Schrijver graphs}. This paper also showed that $\Omega(\log^{\delta}n)$ coloring a $2$-colorable $4$-uniform hypergraph was hard under the assumption P $\neq$ NP. In terms of the largest inapproximability factor, Khot-Saket~\cite{KS14} showed that it is quasi-NP hard to color a $2$-colorable $12$-uniform hypergraph  with $2^{(\log n)^{\Omega(1)}}$ colors. Varma~\cite{Varma15}, in the subsequent work, further reduced the uniformity to $8$ with the same hardness factor as Khot-Saket~\cite{KS14}.

In this work, we make progress on the problem of approximately coloring $2$-colorable $3$-uniform hypergraphs, a problem for which no progress had been made since the result~\cite{DinurRS02}. Following the layered PCP construction of~\cite{DinurGKR05, DinurRS02}, we show that this property of Kneser graphs can be `simulated' by taking a random sampling of {\em Schrijver graph}, induced subgraphs of the Kneser graph, copies. This improves the efficiency of the construction, resulting in the following theorem.
\begin{theorem}
    \label{thm:main_theorem}
    Assuming $\text{NP}\nsubseteq \text{BPTIME}(2^{(\log n)^{\mathcal{O}(1)}})$, given a $2$-colorable $3$-uniform hypergraph $\mathcal{H}$ on $N$ vertices, there is no randomized polynomial time algorithm that colors $\mathcal{H}$ using $\mathcal{O}((\log N)^{\frac{1}{4}})$ colors.
\end{theorem}

    
        
   

The only randomization the reduction uses is in the selection of Schrijver graph copies, specifically in the choice of permutations that generate them. In 2005, Kaplan, Naor, and Reingold~\cite{KaplanNR05} gave an explicit construction of $k$-wise almost independent permutations. This suffices to derandomize our choice, resulting in a deterministic analog of Theorem~\ref{thm:main_theorem}.

\begin{theorem}\label{deterministic_theorem}
    Assume $\text{NP}\nsubseteq \text{DTIME}(2^{\log n^{\mathcal{O}(1)}})$, no polynomial time algorithm can color a $2$-colorable $3$-uniform hypergraph on $N$ vertices with $\mathcal{O}((\log N)^{\frac{1}{8}})$ colors.
\end{theorem}
This theorem improves the result of Dinur, Regev, and Smyth~\cite{DinurRS02} by an exponential factor.

\subsection{Techniques}

Our result follows the reduction of \cite{DinurRS02}. An outline of their reduction is as follows: their outer verifier is an extension of the traditional projection PCP with variable types $X$ and $Y$ that checks if the assignment to the $Y$ variable agrees with the assignment projected onto it by the $X$ variable. They replace each variable with a block of vertices and each test with a collection of hyperedges. However, the standard projection PCP does not suffice: all hyperedges could be satisfied by coloring the $X$ vertices red and the $Y$ vertices blue. To bypass the problem of bipartite-ness, they utilize the layered PCP of \cite{DinurGKR05} (Theorem~\ref{theorem:multiLC}) that introduces variable types beyond $X$ and $Y$. The constructed hypergraph replaces each variable in the layered PCP with a block of vertices corresponding to a certain slice of the Long Code, essentially the Kneser graph with vertex set $\binom{[R]}{s}$ where $R$ is the domain of the variable and $2s + 1 \leq |R|$. Two vertices are adjacent if their corresponding subsets of $R$ are disjoint. It is known that the chromatic number of this graph is $|R| - 2s + 2$. Their construction sets $s \approx \frac{|R|}{2} - \Theta(\log |R|)$, which results in blocks, each with $2^{|R| - \mathcal{O}(\log{|R|})}$ vertices. They show that any attempt to color such a graph using fewer than the chromatic number of colors leaves `many' monochromatic edges, resulting in a decoding. This gives hardness of $\Omega(\sqrt[3]{\log \log n})$ assuming $\text{NP} \nsubseteq \text{DTIME}(2^{\mathrm{poly}(\log n)})$. The size of their reduction (and thus their hardness result) is due to using an entire Kneser graph as the vertex block for each variable. 

The construction in this paper relies on \textit{Schrijver} graphs, a subgraph of the Kneser graph induced by a permutation $\pi$ of the underlying alphabet. See Section~\ref{sec:schrijver} for a formal definition. These sparse subgraphs were already used in the work of Bhangale~\cite{Bhangale18} to improve the hardness of approximate coloring results for hypergraphs that are $4$-uniform. These graphs are much smaller, with $|V(SG_{n,k,\pi})| = n^{\mathcal{O}(1)}$ when $n - 2k = \mathcal{O}(1)$, as compared to the underlying Kneser graphs that have $2^{\Theta(n)}$ vertices. Our construction replaces each variable with a block of vertices corresponding to randomly sampled Schrijver graph copies. Each layer $X_i$ in our reduction has an alphabet $R_i$, and we choose $T_i = |R_i|^{\mathcal{O}(\chi)}$ permutations for our Schrijver blocks at each layer, where $\chi$ is the chromatic number of the Schrijver graph. The starting point of the reduction in~\cite{DinurRS02}, Theorem \ref{theorem:multiLC}, states that for any parameters $\ell,u$ (even as a function of $n$), there exists a reduction from a 3SAT instance of size $n$ to the problem of distinguishing between the completeness and $2^{-\Omega(u)}$ soundness case on an $\ell$-layered Label Cover instance with $n^{\mathcal{O}(u\ell)}$ variables over a range of size $2^{\mathcal{O}(u\ell)}$ that runs in time $n^{\mathcal{O}(u\ell)}$. We set $\ell = 2\chi^2, u = C \cdot \chi$. That means the number of vertices in our hypergraph instance is of size $|V(H)| = n^{\mathcal{O}(u\ell)} \cdot (2^{\mathcal{O}(u\ell)})^\chi = n^{\mathcal{O}(\chi^3)}2^{\mathcal{O}(\chi^4)}$, resulting in an efficient reduction compared to the one in~\cite{DinurRS02}. This ultimately results in the $\mathcal{O}(\log(N)^{1/4})$ hardness factor. 

The reduced size of this construction does not affect the completeness case, but the soundness argument of \cite{DinurRS02} no longer holds, and we have to show that a small number of randomly sampled Schrijver subgraphs is sufficient for the soundness analysis. One key property of the Schrijver graph is that it has the same chromatic number as the Kneser graph from which it is derived, i.e., $\chi(SG_{n,k,\pi}) = \chi(KG_{n,k}) = n - 2k + 2$. However, the Schrijver graph is vertex-critical, so it does not have the property of the Kneser graph mentioned above that results in decoding in ~\cite{DinurRS02}. The core of our proof is to show that randomly sampling a sufficient number of Schrijver graph copies still results in a decoding. The main observation is that the events one needs to care about in the soundness analysis are based on a small collection of small-sized subsets of $R_i$. We show that a small number of randomly sampled Schrijver copies are enough for these events to hold with high probability. This comes from Lemma \ref{mainlemma}, which states that by randomly selecting a family of Schrijver copies, we can still capture the {\em forbidden families} needed for soundness. In particular, consider the set $[n]$ and $\mathcal{A}_{n,d} = \{A \subset [n] : 1 \leq |A| \leq d\}$ where $d$ is small. Take any pairwise disjoint family $\mathcal{Q} \subseteq \mathcal{A}_{n,d}$ and let $\Pi = \{\pi_1, \dots, \pi_T\}$ be independent, uniformly random permutations of $[n]$. Let $\mu$ be the uniform measure on $\binom{[n]}{k}$ and let $\nu_{\Pi}(B) = \mathop{\operatorname{Pr}}_{i\in [T], A \in V(SG_{n,k,\pi_i})}\left[ A= B \right]$. We show via a Hoeffding bounding argument that for the family 
$$\mathcal{F}_\mathcal{Q}=\left\{B\in\binom{[n]}{k}:A\nsubseteq B,\forall A\in \mathcal{Q}\right\}$$
and sufficiently large $T$ we have that with probability at least $1-e^{-\lambda}$ (where $\lambda$ grows with $T$), the following holds simultaneously for {\em every} pairwise disjoint family of sets $\mathcal{Q}$: $\nu_{\Pi}(\mathcal{F}_\mathcal{Q})\geq \rho \implies \mu(\mathcal{F}_\mathcal{Q})\geq\frac{\rho}{2}$. 
Informally, if a forbidden family has noticeable mass among the sampled Schrijver vertices, then it must also have noticeable mass in the full Kneser graph. Combining this statement with the forbidden set counting lemma (Lemma~\ref{countinglemma}) gives exactly the decoding needed for the soundness analysis. This proves the soundness of our reduction and yields Theorem ~\ref{thm:main_theorem}.

We show that this reduction can be derandomized by an explicit, deterministic, choice of the permutations that generate the Schrijver graph using a variant of the almost $k$-wise independent permutations of \cite{KaplanNR05}. We select an $(r,\eta)$-wise uniform distribution over permutations of $[n]$ that has the following two properties: the support has size $e^{\mathcal{O}(r\log n+\log(\frac{1}{\eta}))}$ which can be enumerated in time polynomial in its size, and when we only observe cyclic positions of at most $r$ elements, the distribution behaves almost like a truly uniform random permutation within an error of at most $\eta$. Using the observation that the events we cared about in the above randomized reduction depend on only small subsets of $[n]$, simply replacing the independently sampled permutations of the initial construction with $(r,
\eta)$-wise uniform permutations gives us a deterministic construction, resulting in Theorem \ref{deterministic_theorem}.\\

We note that assuming P$\neq$NP, the best hardness result is by~\cite{DinurRS02}. They showed that for any constant $C>2$, it is NP-hard to color a $2$-colorable $3$-uniform hypergraph with $C$ colors. It is an interesting open question to improve this hardness factor.
\vspace{20pt}

\noindent {\bf Statement of AI use:} The idea of reducing the size of the reduction from~\cite{DinurRS02} using random copies of the Schrijver graph is due to the authors. 
The authors used OpenAI’s ChatGPT Sol 5.6 to assist in developing proofs of some technical lemmas (Lemma~\ref{deterministic_sampler} and Proposition~\ref{negative_association} in particular). The manuscript was written entirely by the authors, with suggestions for stylistic improvements and verification. The authors take full responsibility for its content.

\section{Preliminaries}

\subsection{Label Cover}
\label{sec:lc}

The starting point in our hardness reductions is the \emph{Layered Label Cover} problem, defined next.

\begin{definition}[Layered Label Cover]
An $\ell$-layered Label Cover instance consists of $\ell$ sets of variables $X = \{X_1,\ldots, X_\ell\}$. The range of variables in layer $i$ is denoted by $[R_i]$. Every pair of layers $1\leq i < j \leq \ell$ has a set of constraints $\Phi_{ij}$ between the variables in $X_i$ and $X_j$. The constraint between $x\in X_i$ and $y\in X_j$ is denoted by $\phi_{x\rightarrow y}$. Moreover, every constraint between a pair of variables is a projection constraint -- for every assignment $k \in [R_i]$ to $x$ there is a unique assignment to $y$ that satisfies the constraint $\phi_{x\rightarrow y}$.
\end{definition}

In a Label Cover instance as defined above, for any constraint $\phi_{x \rightarrow y} \in \Phi_{i,j}$, we view it as a function $\phi_{x \rightarrow y}: [R_i] \rightarrow [R_j]$ defined such that for any $k \in [R_i]$, $(k, \phi_{x \rightarrow y}(k))$ satisfies the constraint $\phi_{x \rightarrow y}$. Thus, where there is no ambiguity, we will use $\phi_{x \rightarrow y}$ to denote both the constraint, as well as the function. Moreover, for brevity, we say $x \sim y$, or ``$x$ is a neighbour of $y$'' if $\phi_{x \rightarrow y} \in \Phi_{i,j}$.

\begin{definition}[Weakly dense, \cite{DinurGKR05}]\label{weakly_dense}
  An instance of $\ell$-layered Label Cover is \emph{weakly dense} if the following property holds.
  For any $m$ layers $i_1 < \cdots < i_m$, where $1<m \leq \ell$, and any sequence of variable sets $S_k \subseteq X_{i_k}$ for $k \in [m]$ such that $|S_k| \geq \frac{2}{m}|X_{i_k}|$, we have that there are two sets $S_k$ and $S_{k'}$ such that the number of constraints between $S_k$ and $S_{k'}$ is at least a $\frac{1}{m^2}$ fraction of the total number of constraints between layers $X_{i_k}$ and $X_{i_{k'}}$.
\end{definition}

We have the following $\text{NP}$-hardness result from \cite{DinurGKR05}, \cite{DinurRS02}, which we use as a starting point in proving  Theorem~\ref{thm:main_theorem} and Theorem~\ref{deterministic_theorem}.
\begin{theorem}[\cite{DinurGKR05}, \cite{DinurRS02}]
\label{theorem:multiLC}
For any parameters $\ell,u$, there exists a reduction from an NP-hard problem of size $n$ to the problem of distinguishing between the following two cases in an $\ell$-layered label cover instance $\Phi$ with $n^{\mathcal{O}(u\ell)}$ variables over a range size $2^{\mathcal{O}(u\ell)}$.
\begin{description}
    \item[Completeness.] There exists an assignment satisfying all the tests of $\Phi$.
    
    \item[Soundness.] For every $1 \leq i < j \leq \ell$, no assignment satisfies more than a $2^{-\Omega(u)}$ fraction of the tests in $\Phi_{i,j}$.
\end{description}
Moreover, the reduction runs in time $n^{\mathcal{O}(u\ell)}$ and this layered label cover $\Phi$ also satisfies the weak density property.

\end{theorem}

\subsection{Schrijver Graphs}
\label{sec:schrijver}
We begin by recalling Schrijver graphs. Suppose we have a permutation $\pi$ of $[n]$, we say that two elements $a,b\in [n]$ are $\pi\textit{-adjacent}$ if $a=\pi(i)$ and $b=\pi((i\mod n) + 1)$, for some $1\leq i\leq n$, where indices are taken modulo $n$. For $n>2k$, a set, $S\in \binom{[n]}{k}$ is $\pi\textit{-stable}$ if no pair of elements in $S$ are $\pi$-adjacent. We denote the set of all $\pi\textit{-stable}$ subsets of $[n]$ of size $k$ by $\stab$ and let $\stabsize=|\stab|$. It is a known fact~\cite[Claim 2.1]{DinurRS02} that $\stabsize\leq \binom{n}{n-2k}$ and this quantity is independent of $\pi$.

\begin{definition}
    A Schrijver graph, $\sgraph$ is constructed as follows;
    \begin{itemize}
        \item The vertex set is $\stab$
        \item For $A,B\in \stab$, there is an edge between $A$ and $B$ iff $A\cap B=\emptyset$ i.e, all edges are between disjoint sets.
    \end{itemize}
\end{definition}
It is also well known that the chromatic number of $\sgraph$, $\chromaticnumber=n-2k+2$ \cite{Schrijver78}.\\



Let us recall two combinatorial lemmas from~\cite{DinurRS02} that will be used later in the soundness analysis. The following lemma bounds the size of a pairwise disjoint family of small subsets when a sufficient fraction of $k$ subsets avoids all of them.

\begin{lemma}[\cite{DinurRS02} forbidden-set counting]\label{countinglemma}
There exists an absolute constant $C_0$ such that the following holds. Let $A_1,A_2,\ldots,A_q\subset [n]$ be pairwise disjoint, $|A_i|\leq s$. Let $\mathcal{F}=\{F\in \binom{[n]}{k} \mid \forall A_i\nsubseteq F,\forall i\in[q]\}$. If $\frac{n}{3}\leq k\leq \frac{2n}{3}$ we have;
$$q\leq 3^s\left[\frac{1}{2}\log(n)+C_0-\log\left(\frac{|\mathcal{F}|}{\binom{n}{k}}\right)\right].$$
\end{lemma}

 The next lemma gives us a guarantee of a frequently occurring element; i.e, if a large collection of small sets has only a bounded number of pairwise disjoint members, then some element must belong to a significant fraction of the sets.

\begin{lemma}[\cite{DinurRS02}]\label{packinglemma}
    Let $A_1,A_2\ldots,A_M$ be a collection of $M$ non-empty sets of size at most $s$ such that there are at most $q$ pairwise disjoint sets in the collection. Then there must be an element contained in at least $\frac{M}{qs}$ members of the collection.
\end{lemma}
One  may refer to Claim 4.5 and Claim 4.6 of~\cite{DinurRS02} for the proofs of the above lemmas. We also need the following Hoeffding's inequality.

\begin{lemma}[\cite{Hoeffding63} Hoeffding's theorem]\label{Hoeffding} 
    Let $X_1,\ldots,X_n$ be independent random variables such that $a_i\leq X_i\leq b_i$ almost surely. Consider the sum of these random variables $S_n=\sum_{i=1}^{n}X_i$. Then Hoeffding's theorem states that,
    $$\mathop{\operatorname{Pr}}[S_n -\E[S_n]\geq t]\leq e^{-\left(\frac{2t^2}{\sum_{i=1}^n (b_i-a_i)^2}\right)}.$$
\end{lemma}

The next lemma explains why a small random collection of Schrijver copies is enough for our reduction. The construction in~\cite{DinurRS02} uses all $k$-subsets of the alphabet, and the relevant events in our reduction are exactly the forbidden-set events. After replacing the full Kneser block by sampled Schrijver copies, the following lemma ensures that they still capture the forbidden families needed for soundness.

\begin{lemma}\label{mainlemma}
    Fix $n,k,d$ with $n/3\leq k\leq \lfloor n/2\rfloor$ and fix $\rho \in (0,1)$. Let $\mathcal{A}_{n,d}=\{A\subset [n]: 1\leq |A|\leq d \}$ and $M_{n,d}=|\mathcal{A}_{n,d}|=\sum_{i=1}^{d}\binom{n}{i}$. Let $\Pi=\{\pi_1, \pi_2,\ldots,\pi_T\}$ be independent, uniformly random permutations of $[n]$. For a family $\mathcal{Q}\subseteq \mathcal{A}_{n,d}$, define, 
    $$\mathcal{F}_\mathcal{Q}=\left\{B\in\binom{[n]}{k}:A\nsubseteq B,\forall A\in \mathcal{Q}\right\}.$$
    Let $\mu$ be the uniform measure on $\binom{[n]}{k}$ and let $\nu_{\Pi}$ be a measure on $\binom{[n]}{k}$ defined as follows,
    $$\mbox{for}~ B \in \binom{[n]}{k}, \quad\quad \nu_{\Pi}(B):=\frac{1}{T}\sum_{i=1}^{T}\frac{\mathbf{1}_{B\in \mathrm{Stab}_{\pi_i}(n,k)}}{\stabsize} = \mathop{\operatorname{Pr}}_{\substack{i\in [T]\\ A \in \mathrm{Stab}_{\pi_i}(n,k)}}\left[ A= B \right].$$
    Equivalently, choose $i\in [T]$ uniformly and then choose a uniform set from $\mathrm{Stab}_{\pi_i}(n,k)$.
    If 
    $$T\geq 4\rho^{-2} (M_{n,d} \log 2+\lambda)$$
    then with probability at least $1-e^{-\lambda}$, the following holds simultaneously for every pairwise disjoint family of sets $\mathcal{Q}\subseteq \mathcal{A}_{n,d}$;
    $$\nu_{\Pi}(\mathcal{F}_\mathcal{Q})\geq \rho \implies \mu(\mathcal{F}_\mathcal{Q})\geq\frac{\rho}{2}.$$
\end{lemma}

\begin{proof}
    We want to rule out the following case; $\mu(\mathcal{F}_\mathcal{Q})<\frac{\rho}{2}$ and $\nu_{\Pi}(\mathcal{F}_\mathcal{Q})\geq \rho$. Fix any $B\in \binom{[n]}{k}$, we have
    \begin{align*}
        \mathop{\E}_{\Pi = (\pi_1, \ldots, \pi_T)} \left[\nu_{\Pi}(B)\right] &=   \mathop{\E}_{\Pi = (\pi_1, \ldots, \pi_T)} \left[ \frac{1}{T}\sum_{i=1}^{T}\frac{\mathbf{1}_{B\in \mathrm{Stab}_{\pi_i}(n,k)}}{\stabsize} \right]\\
        & = \frac{1}{T}\sum_{i=1}^{T}\frac{\mathop{\E}_{\pi_i} \left[ \mathbf{1}_{B\in \mathrm{Stab}_{\pi_i}(n,k)}\right]}{\stabsize}.
    \end{align*}
     Now, every $k$-subset $B$ is $\pi$-stable for the same number of permutations. Since every permutation contributes exactly $\stabsize$ stable $k$-sets, averaging over all $B\in \binom{[n]}{k}$ shows that 
    $$\mathop{\E}_{\pi_r} \left[ \mathbf{1}_{B\in \mathrm{Stab}_{\pi_r}(n,k)} \right]= \frac{\stabsize}{\binom{n}{k}}.$$
Therefore, we have,
\begin{equation}\label{eq:uniform}
\mathop{\E}_{\Pi = (\pi_1, \ldots, \pi_T)} \left[\nu_{\Pi}(B)\right]  = \frac{1}{\binom{n}{k}}.
\end{equation}
 In other words, $\mathop{\E}_{\Pi}[\nu_{\Pi}] = \mu$. Pick $\mathcal{Q}\subset \mathcal{A}_{n,d}$. For each $r \in [T]$, let
    \begin{align*}
        Y_{r}&:=\mathop{\operatorname{Pr}}_{S\sim \mathrm{Stab}_{\pi_r}(n,k)}[S\in \mathcal{F}_\mathcal{Q}]\\
        &=\frac{|\mathcal{F}_\mathcal{Q}\cap \mathrm{Stab}_{\pi_r}(n,k)|}{\stabsize}.
    \end{align*} 
 We have $$\nu_{\Pi}(\mathcal{F}_\mathcal{Q})=\frac{1}{T}\sum_{i=1}^TY_{i}.$$ 
 Let us assume that $\mu(\mathcal{F}_\mathcal{Q})<\frac{\rho}{2}$. Now all we need to show is $\nu_{\Pi}(\mathcal{F}_\mathcal{Q})<\rho$. From (\ref{eq:uniform}), we know that $\E_{\Pi}[\nu_{\Pi}(\mathcal{F}_\mathcal{Q})]=\mu(\mathcal{F}_\mathcal{Q})<\frac{\rho}{2}$. If $\nu_{\Pi}(\mathcal{F}_\mathcal{Q})\geq \rho$, then 

$$        \nu_{\Pi}(\mathcal{F}_\mathcal{Q})-\E[\nu_{\Pi}(\mathcal{F}_\mathcal{Q})]\geq \frac{\rho}{2}.$$
Therefore,
    \begin{align*}
        \mathop{\operatorname{Pr}}[\nu_{\Pi}(\mathcal{F}_\mathcal{Q})\geq \rho]&\leq \mathop{\operatorname{Pr}}\left[\nu_{\Pi}(\mathcal{F}_\mathcal{Q})-\E[\nu_{\Pi}(\mathcal{F}_\mathcal{Q})]\geq \frac{\rho}{2}\right]\\
        &\leq \mathop{\operatorname{Pr}}\left[\sum_{i=1}^TY_{i}-\E\left[\sum_{i=1}^TY_{i}\right]\geq\frac{\rho T}{2}\right].
    \end{align*}
We have $0\leq Y_{r}\leq 1$. Hence, by Lemma \ref{Hoeffding}, we have;
\begin{align*}
    \mathop{\operatorname{Pr}}\left[\nu_{\Pi}(\mathcal{F}_\mathcal{Q})\geq \rho\right]&\leq e^{-\frac{\frac{\rho^2 T^2}{2}}{\sum_{i=1}^{T}1 }}\\
    &\leq e^{-\frac{\rho^{2}T}{2}}.
\end{align*}

The number of pairwise disjoint subfamilies of $\mathcal{A}_{n,d}$ is at most $2^{M_{n,d}}$. Thus, the probability that there exists a pairwise disjoint family $\mathcal{Q}$ such that $\mu(\mathcal{F}_\mathcal{Q})<\frac{\rho}{2}$ and $\nu_{\Pi}(\mathcal{F}_\mathcal{Q})\geq \rho$ is at most $2^{M(n,d)}\times e^{-\frac{\rho^2 T}{2}}\leq e^{-\lambda}$. This completes the proof.
\end{proof}

\section{The Reduction}

We now describe the randomized reduction from the layered label cover instance to a $3$ uniform hypergraph. Just as in the original~\cite{DinurRS02} construction, every variable is replaced by a block of vertices indexed by subsets of its alphabet. However, the main difference is that instead of using the entire Kneser block, we use a union of randomly sampled Schrijver copies.

Let $\Phi$ be the layered label cover instance with layers $X_1,X_2,\ldots,X_{\ell}$, alphabets $R_1,R_2,\ldots,R_{\ell}$, projection tests $\test:R_i\rightarrow R_j$ where $x\in X_i$, $y\in X_j$, $i< j$, and $$\ell=2\chi^2\text{, } u=C\cdot {\chi},$$ for some large constant $C$ to be chosen later in the soundness analysis. We write $n_i=|R_i|$, $k_i=\lfloor\frac{n_i-\chi}{2}\rfloor$ and $d_i=n_i-2k_i$. We can observe that $d_i\in \{\chi,\chi+1\}$ and thus $\chi(SG_{n_i,k_i,\pi})=n_i-2k_i+2=d_i+2>\chi$. Thus, every $\chi$ coloring of a sampled Schrijver copy contains a monochromatic edge. We also have $$\frac{n_i}{3}\leq k_i\leq \frac{2n_i}{3}$$ once $u\ell$ is sufficiently large. For each layer $i$, let $$m_i=m(n_i,k_i),\text{ }\rho_i=\frac{1}{\chi m_i},\text{ }M_i=\sum_{a=1}^{\chi+1}\binom{n_i}{a}.$$ Finally, choose 
$$T_i=\lceil 4\rho_i^{-2}(M_i\log2+\log(10\ell))\rceil$$ independent uniformly random permutations 
$$\Pi_i=(\pi_{i,1},\ldots,\pi_{i,T_i})$$ of $R_i.$
\\\\
\textbf{Vertices:} For every variable $x\in X_i$, create a block 
$$V[x]=\{(x,r,S):r\in[T_i], S\in \mathrm{Stab}_{\pi_{i,r}}(n_i,k_i)\}.$$
The vertex set of the hypergraph $H$ is the disjoint union
$$V(H)=\bigcup_{i=1}^{\ell}\bigcup_{x\in X_i}V[x].$$
\\\\
\textbf{Hyperedges:} Consider a projection test $\test:R_i\rightarrow R_j$. Choose two vertices $(x,r,S_1),(x,r,S_2)\in V[x]$ from the same block such that $S_1\cap S_2=\emptyset$. For a vertex $(y,t,U)\in V[y]$, we add a hyperedge between 
$$\{(x,r,S_1),(x,r,S_2),(y,t,U)\}$$
whenever,
$$\test(R_i\setminus (S_1\cup S_2))\subseteq U.$$

This is exactly the \cite{DinurRS02} hyperedge rule except that the source pair and the target set are restricted to sampled Schrijver blocks.

\begin{lemma}[completeness]\label{completeness}
   If $\Phi$ is satisfiable, then the $H$ is 2-colorable for every choice of $\Pi$.  
\end{lemma}
\begin{proof}
    The proof of completeness is the same as the proof in~\cite{DinurRS02}. Let $A$ be a satisfying assignment of $\Phi$. Color a vertex $(x,r,S)$ red whenever $A(x)\in S$ and blue otherwise. Consider the hyperedge $\{(x,r,S_1),(x,r,S_2),(y,t,U)\}$. The hyperedge cannot be all red because the sets $S_1$ and $S_2$ are disjoint. So we only have to look at the case where all three vertices are blue. If all are blue, then this implies $A(x)\in R_i\setminus (S_1\cup S_2)$. So, from our hyperedge construction, we have $\test(A(x))\in U$. But $\test(A(x))=A(y)$ since $A$ satisfies the projection test. This forces $(y,t,U)$ to be colored red. Thus, there is no monochromatic hyperedge.
\end{proof}
\begin{lemma}[soundness]
    If $\Phi$ is not satisfiable, then with a probability of at least 0.9, $H$ is not $\chi$-colorable. 
\end{lemma}
\begin{proof}
    For the sake of contradiction, assume that $H$ admits a $\chi$-coloring. Fix a variable $x\in X_i$. We observed earlier that every sampled Schrijver copy inside $V[x]$ has chromatic number greater than $\chi$, so it contains a monochromatic edge. Record one such edge from each copy and record its color. There is some color $c_x$ that appears in at least $T_i/\chi$ copies. Let $U[x]$ be the vertices in $V[x]$ that are colored $c_x$. Since each copy contributes two such vertices,
    $$\frac{|U[x]|}{|V[x]|}\geq \frac{2T_i/\chi}{T_i\cdot m_i}=\frac{2}{\chi m_i}\geq \rho_i.$$

    Choose one such monochromatic pair $(x,r_i,S_{x,1}), (x,r_i,S_{x,2})$ of color $c_x$ and define it;
    $$B(x)=R_i\setminus(S_{x,1}\cup S_{x,2}).$$
    Then,
    $$|B(x)|=d_i\leq \chi +1.$$

    This $B(x)$ is the list of decoded values. Thus, every variable receives a decoded color and a list of at most $\chi+1$ decoded values. For a layer $X_i$, there is one decoded color that appears on at least $1/\chi$ fraction of its variables. Since $\ell=2\chi^2$, there is a color $\gamma$ that is dominant in at least $2\chi$ layers. By the density property (Definition \ref{weakly_dense}) of the layered label cover, we can find two layers $i\leq j$ and $X\subseteq X_i$,$Y\subseteq X_j$ such that all variables in $X\cup Y$ have decoded color $\gamma$ and at least $1/(4\chi^2)$ fraction of tests are between $X$ and $Y$.

    We will now describe the labeling that we aim to construct. For each $x\in X$, the label $A(x)$ will be chosen uniformly at random from the decoded list $B(x)$. For a variable $y\in Y$, we choose a label $A(y)\in R_j$ that belongs to the largest number of projected lists $\test (B(x))$ where $x\in X$ varies over the neighbors of $y$ in $X$. The rest of the proof will show that there is such a target label that occurs in a large fraction of the projected lists, which in turn implies that many tests between $X$ and $Y$ are satisfied.

    For a fixed $y\in Y$ consider the following multiset;
    $$\mathcal{C}_y=\{\test(B(x)):x\in X\text{ and } \phi_{x\rightarrow y }\text{ exists}\}.$$
    Every member of $\mathcal{C}_y$ has size at most $\chi +1$. 

    \begin{claim}
        There is an element $a_y\in R_j$ that belongs to $$\alpha=\Omega(\chi^{-2}3^{-\chi}(u\ell)^{-1})$$
        fraction of sets in $\mathcal{C}_y$.
    \end{claim}
    \begin{proof}
        Let $A_1,A_2,\ldots,A_q$ be the maximum pairwise disjoint sub-collection of $\mathcal{C}_y$. We will first find a bound on $q$. For every $x\in X$ that has a test to $y$, let $A_x=\test(B(x))$. Observe that no vertex in $U[y]$ can contain any of the $A_i$'s. If there is a vertex $(y,t,U)\in U[y]$ that contains $A_i$ for some $i\in [q]$, then $(x,r_i,S_{x,1}),(x,r_i,S_{x,2}),(y,t,U)$ forms a monochromatic hyperedge. Hence, $U[y]$ avoids every set in $\mathcal{C}_y$. Let,
        $$\mathcal{F}_y=\left\{U\in \binom{R_j}{k_j}:A_i\nsubseteq U, \forall i\in[q]\right\}.$$
 Since every member of $U[y]$ is contained in $\mathcal{F}_y$, we have,
        $$\nu_{\Pi_j}(\mathcal{F}_y)\geq\frac{|U[y]|}{|V[y]|}\geq \rho_j.$$
 Thus, by Lemma \ref{mainlemma}, with a probability of at least $0.9$, we have;
        $$\mu(\mathcal{F}_y)=\frac{|\mathcal{F}_y|}{\binom{n_j}{k_j}}\geq \frac{\rho_j}{2}.$$
        Now apply the forbidden set counting (Lemma \ref{countinglemma}) with $s=\chi +1$ and get
        $$q\leq 3^{\chi+1}\left(\frac{1}{2}\log(n_j)+C_0-\log \left(\frac{|\mathcal{F}_y|}{\binom{n_j}{k_j}}\right)\right)\leq 3^{\chi+1}\left(\frac{1}{2}\log(n_j)+C_0+\log \left(\frac{2}{\rho_j}\right)\right).$$
        Now $\rho_j=\frac{1}{\chi m_j}$ and $m_j\leq \binom{n_j}{d_j}\leq n_j^{\chi+1}$. And thus we get
        $$q\leq \mathcal{O}(\chi3^{\chi}\log n_j).$$
 Since $n_j\leq 2^{\mathcal{O}(u\ell)}$, this becomes
        $$q\leq \mathcal{O}(\chi3^{\chi}u\ell).$$
        Now just apply Lemma~\ref{packinglemma}. It provides a label $a_y\in R_j$ that belongs to at 
        $$\alpha=\frac{1}{q(\chi+1)}=\Omega(\chi^{-2}3^{-\chi}(u\ell)^{-1})$$
        fraction of projected lists in $\mathcal{C}_y.$
   \end{proof}
    Set $A(y)=a_y$ for each $y\in Y$. By construction, this chosen target label belongs to at least $\alpha$ of the projected lists in $\mathcal{C}_y$. Now consider a test $\test$ for which $A(y)\in\test (B(x))$. At least one label in $B(x)$ is mapped to $A(y)$ and since $A(x)$ is chosen uniformly at random from $B(x)$, this test is satisfied with probability at least 
    $$\frac{1}{|B(x)|}\geq \frac{1}{\chi+1}.$$
    Consequently, the expected fraction of tests satisfied between $X$ and $Y$ is at least $\frac{\alpha}{\chi+1}$ and the tests between $X$ and $Y$ is at least $\frac{1}{4\chi^2}$ fraction of the tests between $X_i$ and $X_j$. Thus, in expectation, one obtains a labeling satisfying 
    $$\Omega (\chi^{-5}3^{-\chi}(u\ell)^{-1})$$ fraction of all tests between $i$ and $j$. With $\ell=2\chi^2$ and $u=C\chi$, this lower bound becomes
    $$\Omega (\chi^{-8}3^{-\chi}).$$
    Choosing $C$ sufficiently large will make it greater than the soundness threshold of $2^{-\Omega(u)}$ which is a contradiction. Thus, our assumption is incorrect, and $H$ is not $\chi$-colorable. Since this holds with probability greater than 0.9, the soundness follows.
\end{proof}

We now prove Theorem \ref{thm:main_theorem} by understanding the parameters in the above reduction.\\

\noindent {\bf Proof of Theorem \ref{thm:main_theorem}:} Fix a 3SAT instance of size $n_0$. Take $\chi$ to be a sufficiently small constant multiple of $\log(n_0)$. For the parameters $u=C\cdot \chi$ and $\ell = 2\chi^2$, using Theorem~\ref{theorem:multiLC}, we first reduce a 3SAT instance of size $n_0$ to a layered label cover instance. We then reduce the layered label cover instance to a $3$-uniform hypergraph using the above reduction.

Recall that $d_i\leq \chi+1$, so $m_i\leq\binom{n_i}{d_i}\leq n_i^{\chi+1}$ and $M_i\leq (\chi+1)n_i^{\chi+1}$. Therefore, 
$$T_i=\mathcal{O}(\rho^{-2}(M_i+\log \ell))\leq n_i^{\mathcal{O}(\chi)}.$$

For each variable $x\in X_i$, the corresponding block contains one copy of $\mathrm{Stab}_{\pi_{i,r}}(n_i,k_i)$ for each sampled permutation. Thus,
$$|V[x]|=T_im_i\leq n_i^{\mathcal{O}(\chi)}.$$

Summing over all variables and using the fact that the layered label cover contains $n_0^{\mathcal{O}(u\ell)}$ variables and has an alphabet size of at most $2^{\mathcal{O}(u\ell)}$, we get,
$$N:=|V(H)|\leq n_0^{\mathcal{O}(\chi^3)}2^{\mathcal{O}(\chi^4)}.$$
We have $\log(|V(H)|)=\mathcal{O}(\chi^4)$. So,
$$\chi=\Omega((\log|V(H)|)^{1/4}).$$

 Note also that the overall reduction runs in randomized time $n_0^{\mathcal{O}(u\ell)} = 2^{(\log n_0)^{\mathcal{O}(1)}}$. Thus, from the completeness and soundness of the reduction, any randomized polynomial time coloring algorithm that colors a $2$-colorable $3$-uniform hypergraph on $N$ vertices with fewer than $\mathcal{O}(\log(N)^{1/4})$ colors would put 3SAT in $\text{BPTIME}(2^{(\log n)^{\mathcal{O}(1)}})$ and thus, contradicting our assumption that $\text{NP}\nsubseteq \text{BPTIME}(2^{(\log n)^{\mathcal{O}(1)}})$. This completes the proof of Theorem \ref{thm:main_theorem}.\qed

\section{Derandomization}

The only random choice in the above reduction is the collection of permutations used to form the sampled Schrijver copies. The randomness is used only in Lemma \ref{mainlemma} to guarantee that the sampled copies satisfy the forbidden set events needed for the soundness proof. This suggests replacing the random family of permutations with an explicit family that still satisfies the properties needed for soundness. We therefore look for a deterministic family of permutations that behaves like a uniformly random permutation. An appropriate notion for this is an almost-wise uniform family of permutations, which we define next.

\begin{definition}\label{def:derand_per}
    A distribution $\mathcal{D}$ over permutations of $[n]$ is called $(r,\eta)$-wise uniform if, for every set $B\subseteq [n]$ with $|B|\leq r$ the distribution of the cyclic positions of the elements of $B$ under $\pi \sim \mathcal{D}$ is within total variation distance $\eta$ of the corresponding distribution under a uniformly random permutation.
\end{definition}

Here, the total variation distance measures how different two probability distributions are. For two distributions $P$ and $Q$ on the same finite sample space $\Omega$, it is defined as
$$d_{TV}(P,Q)=\frac{1}{2}\sum_{\omega\in \Omega}|P(\omega)-Q(\omega)|.$$
Thus, an $(r,\eta)$-wise uniform distribution behaves almost like a truly uniform random permutation when we only observe the cyclic positions of at most $r$ elements, with an error of at most $\eta$.

\begin{theorem}[Explicit almost-wise uniform permutations.] For every $\eta,r$ and $\eta>0$, there is an explicit $(r,\eta)$-wise uniform distribution over permutations of $[n]$ whose support has size at most 
$$e^{\mathcal{O}(r\log n+\log(\frac{1}{\eta}))}$$
Moreover, the support can be enumerated in time polynomial in its size.  
\end{theorem}

This follows from standard constructions of almost $k$-wise independent permutations, for example, the construction of  Kaplan, Naor, and Reingold \cite{KaplanNR05}. In the randomized proof, we used Lemma \ref{mainlemma} to show that if a forbidden family has a large measure among the sampled Schrijver copies, then it also contains a large fraction of all $k$-subsets. This allowed us to apply Lemma \ref{countinglemma} and bound the number of pairwise disjoint sets. In the deterministic proof, we will make use of the following lemma and obtain this bound directly.

\begin{lemma}\label{deterministic_sampler}
    There are absolute constants $C_2,C_3,C_4$ such that the following holds. Let $\frac{n}{3}\leq k \leq \lfloor\frac{n}{2}\rfloor$ let $s\geq 1$ assume that $n\geq 12s$. Let $\rho\in (0,\frac{1}{2})$ and let h be an even integer satisfying $h\geq C_2\log (\frac{1}{\rho})$. Set 
    $$r=sh,\text{   } M=\sum_{a=1}^{s}\binom{n}{a}.$$
    Let $\mathcal{D}$ be an $(r,\eta)$-wise uniform distribution over permutations of $[n]$ with $\eta \leq \frac{\rho}{4}\left(\frac{C_3h}{2M} \right)^h$. 
    For every pairwise disjoint family $A_1.\cdots, A_q\subseteq [n]$. $1\leq |A_i|\leq s$, define 
    $$\mathcal{F}=\left\{ S\in\binom{[n]}{k}: A_i \nsubseteq  S ~\forall i\in [q]\right\}.$$ 
    Let $\Pi$ be the support of $\mathcal{D}$, with multiplicity according to uniform seeds, and define $v_{\Pi}$ by choosing a seed uniformly and then a uniform stable $k$-set for the resulting permutation, i.e., 
    $$\mbox{for}~ B \in \binom{[n]}{k}, \quad\quad \nu_{\Pi}(B):=\frac{1}{|\Pi|}\sum_{\pi_i\in\Pi}\frac{\mathbf{1}_{B\in \mathrm{Stab}_{\pi_i}(n,k)}}{\stabsize}.$$
    If $$v_{\Pi}(\mathcal{F})\geq \rho,$$ then $$q\leq C_44^sh.$$
\end{lemma}

\begin{proof}
    For contradiction, assume that $q>C_44^sh$, where $C_4$ will be chosen large enough. Let $S$ be a uniformly random $k$-subset of $[n]$ and let $\mu$ be a uniform measure on $\binom{[n]}{k}$. For each $i\in[q]$, let $Y_i=\mathbf{1}[A_i\subseteq S]$ and $Z=\sum_{i=1}^qY_i$. Note that $S\in \mathcal{F}$ iff $Z=0$. Therefore, to show that $\mathcal{F}$ has a small measure, it is enough to show that $Z=0$ is less likely. Since $|A_i|\leq s$, $n\geq 12s$ and $k\geq\frac{n}{3}$, for every $i$, we have
    $$\mathop{\operatorname{Pr}}[A_i\subseteq S]=\frac{(k)_{|A_i|}}{(n)_{|A_i|}}\geq 4^{-s}.$$
       Here $(k)_r=k\cdot (k-1)\cdots(k-r+1)$ denotes the falling factorial. Let $\lambda= \E_{\mu}[Z]$, where $\mu$ is the uniform measure. We get $\lambda\geq q4^{-s}\geq C_4h$. 

    \begin{claim}
        The variables $Y_1,\ldots,Y_q$ are negatively associated.
    \end{claim}
    The proof is given in Appendix, Proposition \ref{negative_association}. Since the $Y_i$'s are negatively associated Bernoulli variables, the usual Chernoff-Hoeffding tail bounds apply to $Z=\sum_{i=1}^qY_i$ (see Proposition~\ref{prop:chernoff_neg}). Hence,
    $$\mathop{\operatorname{Pr}}[|Z-\lambda|\geq t]\leq 2\exp \left(-c\min\left\{\frac{t^2}{\lambda},t\right\}\right)$$ for some absolute constant $c$. Using the standard tail-to-moment identity,
    $$\E_{\mu}\left[|Z-\lambda|^h\right]=h\int_{0}^{\infty}t^{h-1}\mathop{\operatorname{Pr}}(|Z-\lambda|\geq t)dt,$$
    and since $\lambda\geq C_4h$, we get for an even $h$,
    $$\E_{\mu}\left[(Z-\lambda)]^h\right]\leq(Ch\lambda)^{\frac{h}{2}}.$$
    
    Observe that the expansion of $(Z-\lambda)^h$ has terms consisting of a coefficient times a product of several $Y_i$'s. These products involve an event of the form 
    $$\bigcap_{i\in I}\{A_i\subseteq S\}=\left\{\bigcup_{i\in I}A_i\subseteq S\right\}$$ for some $I\subseteq[q]$ with $|I|\leq h$. Since the $A_i's$ are pairwise disjoint, the union consists of at most $sh=r$ elements. Thus, the $(r,\eta)$-wise uniformity of $\mathcal{D}$ implies that the expectation of each monomial differs from its expectation under a truly uniform permutation by at most $\eta$. The total absolute coefficient mass in the expansion is at most $(\lambda +q )^h\leq (2q)^h\leq (2M)^h$. Therefore, 
    $$\left |\E_{\nu_{\Pi}}[(Z-\lambda)^h]-\E_{\mu}[(Z-\lambda)^h)]\right |\leq \eta(2M)^h\leq \frac{\rho}{4}(C_3h)^h.$$
    Choosing the constants appropriately and using $\lambda>C_4 h$ gives, 
    $$\E_{\nu_{\Pi}}[(Z-\lambda)^h]\leq (Ch\lambda)^{\frac{h}{2}}+\frac{\rho}{4}\lambda^h.$$
    Now $S\in \mathcal{F}$ exactly when $Z=0$ and if $Z=0$, then $|Z-\lambda|=\lambda$. Thus 
    $$\nu_{\Pi}(\mathcal{F})=\mathop{\operatorname{Pr}}_{\nu_{\Pi}}(Z=0)\leq\mathop{\operatorname{Pr}}_{\nu_{\Pi}}(|Z-\lambda|=\lambda)\leq \mathop{\operatorname{Pr}}_{\nu_{\Pi}}(|Z-\lambda|\geq\lambda).$$ Therefore Markov's inequality gives, 
    $$\nu_{\Pi}(\mathcal{F})\leq \mathop{\operatorname{Pr}}_{\nu_{\Pi}}[|Z-\lambda|\geq\lambda]\leq\frac{\E_{\nu_{\Pi}}[(Z-\lambda)^h]}{\lambda^h}\leq \left(\frac{Ch}{\lambda}\right)^{\frac{h}{2}}+\frac{\rho}{4}.$$
    Earlier we showed that $\lambda\geq C_4h$. Thus $\left(\frac{Ch}{\lambda}\right)^{\frac{h}{2}}\leq\left(\frac{C}{C_4}\right)^{\frac{h}{2}}$. Taking $C_4$ sufficiently large, we get $\left(\frac{C}{C_4}\right)^{\frac{h}{2}}\leq \frac{1}{4}^{\frac{h}{2}}=e^{-c'h}$ for an absolute constant $c'> 0$. Since $h\geq C_2\log(\frac{1}{\rho})$, taking $C_2$ large enough makes this term at most $\frac{\rho}{2}$. Hence $\nu_{\Pi}(\mathcal{F})<\rho$, contradicting the assumption. Therefore $q\leq C_44^sh.$
    
\end{proof}

\subsection{The deterministic reduction}

We now return to the layered label cover instance and replace  independently sampled permutations with explicit $(r,\eta)$-wise uniform permutations. The hypergraph itself remains unchanged. 

For a layer $i$, recall,
$$n_i=|R_i|,\text{   }  k_i=\left\lfloor \frac{n_i-\chi}{2}\right\rfloor, \text{   }m_i=m(n_i,k_i),\text{   }\rho_i=\frac{1}{\chi m_i}. $$
Set $s=\chi+1$ and let $h_i$ be the least even integer satisfying $h_i\geq C_2 \log(\frac{1}{\rho_i})$ and define, 

$$r_i=sh_i, \textbf{   }M_i=\sum_{a=1}^s\binom{n_i}{a}\text{    },\eta_i=\frac{\rho_i}{4}\left(\frac{C_3h_i}{2M_i}\right)^{h_i}.$$

\noindent\textbf{{Vertices:}} choose an explicit $(r_i,\eta)$-wise uniform distribution $\mathcal{D}_i$ on permutations of $R_i$ and let $\Omega_i$ be its uniform seed set. If the seed $\omega\in \Omega_i$ produces the permutation $\pi_{i,\omega}$, then for each variable $x\in X_i$, we create
$$V[x]=\{(x,\omega,S) :\omega\in \Omega_i,S\in \mathrm{Stab}_{\pi_{i,\omega}}(n_i,k_i)\}.$$
Thus, every seed gives one Schrijver copy, and repeated permutations produced by different seeds are kept as repeated copies. 

\noindent{\textbf{Hyperedges:}} the hyperedge rule is unchanged. For every projection $\test:R_i\rightarrow R_j $, take two disjoint vertices $(x,\omega,S_1)$ and $(x,\omega,S_2)$ from the same seed copy and a target vertex $(y,\tau,U)$. Add an edge
$$\{(x,\omega,S_1),(x,\omega,S_2),(y,\tau,U)\}$$
whenever
$$\test(R_i\setminus (S_1\cup S_2))\subseteq U.$$

The completeness is same as Lemma \ref{completeness}.

\begin{lemma}[soundness]\label{deterministic_soundness}
    If $\phi$ is not satisfiable, then $H$ is not $\chi$ colorable.
\end{lemma}

\begin{proof}
    For contradiction let us assume that $H$ is $\chi$ colorable. The local decoding argument is unchanged. Since every seed copy has chromatic number greater than $\chi$, the corresponding Schrijver graph contains a monochromatic edge. Just like in the randomized soundness there is some color $c_x$ that appears in more than $\frac{1}{\chi}$ fraction of the seed copies and $U[x]\subseteq V[x]$ be the vertices in block $V[x]$ that receive the color $c_x$. We have $\frac{U[x]}{V[x]}\geq \rho_i$. Pick the decoded list $B[x]$ as before. We have $|B[x]|\leq \chi+1$. 

    The layer selection argument is also unchanged. We obtain two layers $i\leq j$, subsets $X\subseteq X_i$ and $Y\subseteq X_j$, and a common decoded color $\gamma$, such that the test between $X$ and $Y$ form an $\Omega (\chi^{-2})$ fraction od all tests between the two layers. 

    Fix $y\in Y$ with at least one incident test from $X$, define the multiset,
    $$\mathcal{C}_y =\{\test(B(x)):x\in X \text{ and } \test \text{ exists}\}.$$

    Just as in the randomized proof, every member of $\mathcal{C}_y$ is non empty and has size at most $s=\chi+1$ and every vertex of $U[y]$ avoids every member of $\mathcal{C}_y$. 

    Let $A_1,\ldots,A_q$ be a maximum pairwise disjoint sub collection of $\mathcal{C}_y$, and define
    $$\mathcal{F}_y=\left\{ U\in \binom{R_j}{k_j}:A_r \nsubseteq  U,\forall r\in q\right\}.$$
    Since every member of $U[y]$ is contained in $\mathcal{F}_y$, 
    $$\nu_{\Pi_j}(\mathcal{F}_y)\geq\frac{|U[y]|}{|V[y]|}\geq \rho_j.$$
    Applying Lemma \ref{deterministic_sampler} gives, 
    $$q\leq \mathcal{O}(4^{\chi}h_j).$$
    Moreover, 
    $$h_j=\mathcal{O}\left(\log \left(\frac{1}{\rho_j}\right)\right)=\mathcal{O}\left(\chi\log n_j\right)=\mathcal{O}\left(\chi u\ell\right),$$
    so,
    $$q\leq\mathcal{O}\left(\chi 4^{\chi}u\ell\right).$$
    Thus, Lemma~\ref{packinglemma} gives us a label $a_y$ contained in at least

    $$\alpha_{det}=\Omega\left( \chi^{-2}4^{-\chi}(u\ell)^{-1}\right)$$
    of the projected lists in $\mathcal{C}_y$.

    Set $A(y)=a_y$ as before; choose $A(x)$ uniformly from $B(x)$, and the corresponding test is satisfied with probability at least $\frac{1}{\chi+1}$. Combining this with the $\Omega(\chi^{-2})$ tests between $X_i$ and $X_j$, The expected fraction of satisfied tests between layers $i$ and $j$ is 
    $$\Omega\left( \chi^{-3}\alpha_{det}\right)=\Omega\left(\chi^{-5}4^{-\chi}(u\ell)^{-1}\right).$$
    With $\ell=2\chi^2$ and $u=C\chi$, this becomes
    $$\Omega\left(\chi^{-8}4^{-\chi}\right).$$
    Choosing $C$ sufficiently large makes this larger than the soundness threshold $2^{-\Omega(u)}$, which is a contradiction. Hence, the hypergraph is not $\chi$-colorable. This completes the proof of Lemma \ref{deterministic_soundness}.
\end{proof}

We now prove Theorem~\ref{deterministic_theorem}.\\

\noindent{\bf Proof of Theorem \ref{deterministic_theorem}:} First, we bound the number of seed copies introduced by de-randomization. Recall $m_i\leq n_i^{\mathcal{O}(\chi)}$, $\rho_i^{-1}=\chi m_i$. Therefore, $\log(\frac{1}{\rho_i})=\mathcal{O}(\chi\log n_i)$, and consequently,
    $$h_i=\mathcal{O}\left(\chi \log n_i\right), \text{   }r_i=\left(\chi+1\right)h_i=\mathcal{O}\left(\chi^2\log n_i\right). $$
    Also, 
    $$M_i=\sum_{a=1}^{\chi+1}\binom{n_i}{a}\leq n_i^{\mathcal{O}(\chi)},$$
    so, 
    $$\log M_i=\mathcal{O}\left(\chi\log n_i\right).$$
    From
    $$\eta_i=\frac{\rho_i}{4}\left(\frac{C_3h_i}{2M_i}\right)^{h_i},$$
    we obtain, 
    \begin{align*}
    \log \left(\frac{1}{\eta_i}\right)&=\mathcal{O}\left(\log\left(\frac{1}{\rho_i}\right)+h_i\log\left(\frac{M_i}{h_i}\right)\right)
    \\
    &=\mathcal{O}\left(\chi^2\left(\log n_i\right)^2\right).
    \end{align*}
    The explicit almost-wise uniform family therefore has a seed size at most 
    $$\exp\left(\mathcal{O}\left(r_i\log n_i+\log\left(\frac{1}{\eta_i}\right)\right)\right)=\exp\left(\mathcal{O}\left(\chi^2\left(\log n_i\right)^2\right)\right).$$
    Since $\log n_i=\mathcal{O}(u\ell)=\mathcal{O}(\chi^3)$, substituting this into the seed size exponent yields,
    $$|\Omega_i|\leq 2^{\mathcal{O}(\chi^8)}.$$
    Each seed contributes one stable set copy of size
    $$m_i\leq n_i^{\mathcal{O}(\chi)}=2^{\mathcal{O}(\chi^4)},$$
    so each block has size $2^{\mathcal{O}(\chi^8)}$. Starting with the 3SAT instance of size $n_0$, the final hypergraph therefore has, 
    $$|V(H)|\leq n_0^{\mathcal{O}(u\ell)}2^{\mathcal{O}(\chi^8)}.$$
    Taking $\chi=\Theta(\log n_0)$ yields
    $$\log |V(H)|=\mathcal{O}((\log n_0)^8),\qquad \chi=\Omega\left(\left(\log |V(H)|\right)^{\frac{1}{8}}\right)$$
     and the overall reduction runs in deterministic time $2^{\mathcal{O}((\log n)^{\mathcal{O}(1)})}$. This proves Theorem~\ref{deterministic_theorem}.\qed

\bibliographystyle{alpha}
\bibliography{references}

\newcommand{\etalchar}[1]{$^{#1}$}
\begin{thebibliography}{GHH{\etalchar{+}}14}

\bibitem[Bha18]{Bhangale18}
Amey Bhangale.
\newblock {NP-Hardness of Coloring 2-Colorable Hypergraph with
  Poly-Logarithmically Many Colors}.
\newblock In {\em 45th International Colloquium on Automata, Languages, and
  Programming (ICALP 2018)}, volume 107, pages 15:1--15:11, Dagstuhl, Germany,
  2018. Schloss Dagstuhl -- Leibniz-Zentrum f{\"u}r Informatik.

\bibitem[DGKR05]{DinurGKR05}
Irit Dinur, Venkatesan Guruswami, Subhash Khot, and Oded Regev.
\newblock A new multilayered pcp and the hardness of hypergraph vertex cover.
\newblock {\em SIAM Journal on Computing}, 34(5):1129--1146, 2005.

\bibitem[DR96]{DubhashiR96}
Devdatt~P. Dubhashi and Desh Ranjan.
\newblock Balls and bins: A study in negative dependence.
\newblock {\em BRICS Report Series}, 3(25), Jan. 1996.

\bibitem[DRS05]{DinurRS02}
Irit Dinur, Oded Regev, and Clifford Smyth.
\newblock The hardness of 3-uniform hypergraph coloring.
\newblock {\em Combinatorica}, 25(5):519--535, 2005.

\bibitem[GHH{\etalchar{+}}14]{GHHSV17}
Venkatesan Guruswami, Prahladh Harsha, Johan H{\aa}stad, Srikanth Srinivasan,
  and Girish Varma.
\newblock {Super-polylogarithmic Hypergraph Coloring Hardness via Low-degree
  Long Codes}.
\newblock In {\em Proceedings of the Forty-sixth Annual ACM Symposium on Theory
  of Computing}, STOC '14, pages 614--623, New York, NY, USA, 2014. ACM.

\bibitem[GHS02]{GHS02}
Venkatesan Guruswami, Johan H{\aa}stad, and Madhu Sudan.
\newblock Hardness of approximate hypergraph coloring.
\newblock {\em SIAM Journal on Computing}, 31(6):1663--1686, 2002.

\bibitem[Hoe63]{Hoeffding63}
Wassily Hoeffding.
\newblock Probability inequalities for sums of bounded random variables.
\newblock {\em Journal of the American Statistical Association},
  58(301):13--30, 1963.

\bibitem[KNR05]{KaplanNR05}
Eyal Kaplan, Moni Naor, and Omer Reingold.
\newblock Derandomized constructions of k-wise (almost) independent
  permutations.
\newblock {\em Algorithmica}, 55:113--133, 2005.

\bibitem[KNS01]{KrivNS01}
Michael Krivelevich, Ram Nathaniel, and Benny Sudakov.
\newblock Approximating coloring and maximum independent sets in 3-uniform
  hypergraphs.
\newblock {\em Journal of Algorithms}, 41(1):99--113, 2001.

\bibitem[KS14]{KS14}
Subhash Khot and Rishi Saket.
\newblock {Hardness of Coloring 2-Colorable 12-Uniform Hypergraphs with exp
  (log\^{}$\{$Omega (1)$\}$ n) Colors}.
\newblock In {\em Foundations of Computer Science (FOCS), 2014 IEEE 55th Annual
  Symposium on}, pages 206--215. IEEE, 2014.

\bibitem[Sak14]{S14}
Rishi Saket.
\newblock {Hardness of finding independent sets in 2-colorable hypergraphs and
  of satisfiable CSPs}.
\newblock In {\em IEEE 29th Conference on Computational Complexity (CCC)},
  pages 78--89, 2014.

\bibitem[Sch78]{Schrijver78}
A.~Schrijver.
\newblock Vertex-critical subgraphs of kneser-graphs.
\newblock {\em Nieuw Archief voor Wiskunde}, 26(3):454--461, 1978.

\bibitem[Var15]{Varma15}
Girish Varma.
\newblock Reducing uniformity in {K}hot-{S}aket hypergraph coloring hardness
  reductions.
\newblock {\em Chicago Journal OF Theoretical Computer Science}, 3:1--8, 2015.

\end{thebibliography}
\appendix
\section{Negatively Associated Random Variables}

We briefly collect facts about negatively associated random variables that are used in the deterministic sampler argument.

\begin{definition}[Negative Association \cite{DubhashiR96}]
    Let $X=(X_1,\ldots,X_n)$ be a vector of random variables. They are \textbf{negatively associated} $((-A)$ property$)$ if for every two disjoint index sets, $I,J\subseteq [n]$,
    $$\E\left[f(X_i,i\in I)g(X_j,j\in J)\right]\leq \E\left[f(X_i,i\in I)\right]\E\left[g(X_j,j\in J)\right]$$
    for all functions $f:\mathbb{R}^{|I|}\rightarrow \mathbb{R}$, $g:\mathbb{R}^{|J|}\rightarrow \mathbb{R}$ and both are coordinate wise non decreasing. Equivalently,
    $$Cov(f(X_i:i\in I),g(X_j:j\in J))\leq 0.$$
    
\end{definition}

The following proposition from~\cite{DubhashiR96} gives the concentration result that we need for sum of negatively associated variables.

\begin{proposition}[\cite{DubhashiR96}]
\label{prop:chernoff_neg}
    Let $X_1,\ldots,X_n$ be negatively associated random variables taking values in $[0,1]$, and let $Z=\sum_{i=1}^n X_i$ with $\mu=\E[Z]$. Then the standard Chernoff-Hoeffding bounds are applicable to $Z$. Consequently, for every $t>0$
    $$\mathop{\operatorname{Pr}}[|Z-\mu|\geq t]\leq 2\exp\left(-c \min \left\{ \frac{t^2}{\mu},t\right\}\right)$$ for some absolute constant $c>0$.
\end{proposition}

We prove here the negative association property used in the deterministic sampler argument. Recall that $S$ is choose uniformly from $\binom{[n]}{k}$, the sets $A_1,\ldots,A_q$ are pairwise disjoint and $Y_i=\mathbf{1}[A_i\subseteq S].$

\begin{proposition}\label{negative_association}
    The variables $Y_1,\ldots,Y_q$ are negatively associated.
\end{proposition}
\begin{proof}
    For each $a\in [n]$, define the indicator,
    $$X_a=\mathbf{1}[a\in S].$$
    Observe that $Y_i=\prod_{a_i\in A_i}X_{a_i}$.
    \begin{claim}
        $Y_1,\ldots,Y_q$ are negatively associated if $X_1,\ldots,X_n$ are negatively associated 
    \end{claim}
    \begin{proof}
        If $X_1,\ldots,X_n$ satisfy $(-A)$ condition then for every $I,J\in [q], I\cap J=\emptyset$ and for every non decreasing functions $f:\mathbb{R}^{|I|}\rightarrow \mathbb{R}$ and $g:\mathbb{R}^{|J|}\rightarrow \mathbb{R}$, 
        $$\E\left[f(X_i,i\in I)g(X_j,j\in J)\right]\leq \E\left[f(X_i,i\in I)\right]\cdot\E\left[g(X_j,j\in J)\right].$$
        Let $U=\cup_{i\in I}A_i$ and $V=\cup_{j\in J}A_j$. $U\cap V=\emptyset$ since $A_1,\ldots,A_q$ are pairwise disjoint. Consider the following,
        $$f(Y_i)_{i\in I}=f\left(\prod_{a_i\in A_i}X_{a_i}\right)_{i\in I}=F((X_a)_{a\in U}).$$
        Observe that $F$ is still a non decreasing function on $(X_a)_{a\in U}$ as changing a $X_a$ from $0$ to $1$ can never decrease the value of $Y_i$. Define $g(Y_j)_{j\in J}=G(X_b)_{b\in V}$ in a similar manner. We have,
        \begin{align*}
            \E\left[f(Y_i)_{i\in I}g(Y_j)_{j\in J}\right] &=\E\left[f\left(\prod_{a_i\in A_i}X_{a_i}\right)_{i\in I}\cdot g\left(\prod_{a_j\in A_j}X_{a_j}\right)_{j\in J}\right]\\
            &=\E\left[F\left(X_a\right)_{a\in U} \cdot G\left(X_b\right)_{b\in V}\right]\\
            &\leq \E\left[F\left(X_{a}\right)_{a\in U}\right]\cdot\E\left[G\left(X_{b}\right)_{b\in V}\right]\\
            &= \E\left[f\left(\prod_{a_i\in A_i}X_{a_i}\right)_{i\in I}\right]\cdot\E\left[g\left(\prod_{a_j\in A_j}X_{a_j}\right)_{j\in J}\right]\\
            &=\E\left[ f(Y_i)_{i\in I}\right]\cdot   \E\left[g(Y_j)_{j\in J}\right].
        \end{align*}

    \end{proof}

   So it is enough to show that $X_1,\ldots, X_n$ are negatively associated. Let $I$ and $J$ be two disjoint subsets of $[n]$ and $f:\mathbb{R}^{|I|}\rightarrow \mathbb{R}$, $g:\mathbb{R}^{|J|}\rightarrow \mathbb{R}$  are two non decreasing functions. Suppose that $|S\cap I|=r$. This leaves only $k-r$ elements that are potentially in $J\cap S$. Thus as $r$ increases, it leaves fewer choices available for $J\cap S$. 
    Let $T_r$ be a uniformly chosen set from $\binom{[n]\setminus I}{k-r}$ and $T_{r+1}$ be obtained after removing an element uniformly at random from $T_r$. $T_{r+1}$ is a uniformly chosen set from $\binom{[n]\setminus I}{k-r-1}$ and $T_{r+1}\cap J\subseteq T_r\cap J$. Let $\mathcal{D}$ be the distribution on $|S\cap I|$ and define the following functions, 
    \begin{align*}
        F(r)&=\E\left[f(X_i)_{i\in I}|r)\right]\\
        G(r)&=\E\left[g(X_j)_{j\in J}|r)\right].
    \end{align*}
    Observe that $F$ is non decreasing on $r$ and since $T_{r+1}\cap J\subseteq T_r\cap J$, $G$ is non increasing on $r$. 
    \begin{claim}\label{claim:FG}
    If $F$ is non decreasing and $G$ is non increasing on $r$, then,
        $$\E_{r\sim \mathcal{D}}\left[F(r)G(r)\right]\leq \E_{r\sim \mathcal{D}}\left[F(r)\right]\E_{r\sim \mathcal{D}}\left[G(r)\right].$$
    \end{claim}
    \begin{proof}
        Let $r$ and $r'$ be two independent variables. Since $F$ is non decreasing and $G$ is non increasing, we have
        $$(F(r)-F(r'))\cdot (G(r)-G(r'))\leq 0.$$
        Therefore,
        \begin{align*}
            \E_{r,r'\sim \mathcal{D}}\left[(F(r)-F(r'))\cdot (G(r)-G(r'))\right] &= \E_{r\sim \mathcal{D}}[F(r)\cdot G(r)]-\E_{r,r'\sim \mathcal{D}}[F(r)\cdot G(r')]\\&\quad\quad -\E_{r,r'\sim \mathcal{D}}[F(r')\cdot G(r)]+\E_{r'\sim \mathcal{D}}[F(r')\cdot G(R')]\\
            &\leq 0.
        \end{align*}
        Now $\E_{r,r'\sim \mathcal{D}}[F(r')G(r)]=\E_{r,r'\sim \mathcal{D}}[F(r)G(r')]$ and $\E_{r\sim \mathcal{D}}[F(r)G(r)]=\E_{r'\sim \mathcal{D}}[F(r')G(r')]$. 
        Therefore
        $$2\E_{r\sim \mathcal{D}}[F(r)G(r)]-2\E_{r,r'\sim \mathcal{D}}[F(r)\cdot G(r')] \leq 0$$
        and since $r$ and $r'$ are chosen independently, this implies,
        \begin{align*}
        \E_{r'\sim \mathcal{D}}[F(r)G(r)]&\leq \E_{r\sim \mathcal{D}}[F(r)]\cdot \E_{r'\sim \mathcal{D}}[G[r']\\
        &=\E_{r\sim \mathcal{D}}[F(r)]\cdot \E_{r\sim \mathcal{D}}[G(r)].
        \end{align*}
    \end{proof}
Note that $(X_i)_{i\in I}$ and $(X_j)_{j\in J}$ are independent when conditioned on $r$. Thus, 
    \begin{align*}
        \E_{r\sim \mathcal{D}}\left[F(r)G(r)\right]&=\E_{r\sim \mathcal{D}}\left[\E\left[f((X_i){i\in I})|r\right]\cdot \E\left[g((X_j)_{j\in J})|r\right]\right]\\
        &=\E\left[(f(X_i)_{i\in I}).(g(Y_j)_{j\in J})\right]
    \end{align*}
    and, 
    $$\E_{r\sim \mathcal{D}}[F(r)].\E[G(r)]=\E\left[f(X_i)_{i\in I}\right]\cdot \E\left[g(X_j)_{j\in J}\right].$$
    Therefore, from Claim~\ref{claim:FG}, we get
    $$\E\left[(f(X_i)_{i\in I}).(g(X_j)_{j\in J})\right]\leq \E\left[f(X_i)_{i\in I}\right]\cdot \E\left[g(X_j)_{j\in J}\right].$$
    This proves $X_1,\ldots,X_n$ are negatively associated.
\end{proof}
\end{document}